\documentclass[showpacs, pra,twocolumn,preprintnumbers ,amsmath, amssymb, superscriptaddress, aps]{revtex4-2}

\usepackage[utf8]{inputenc}
\DeclareUnicodeCharacter{202F}{\,}
\usepackage{color}
\usepackage{amsmath,amssymb}
\usepackage{pifont}
\usepackage{amssymb}  
\usepackage{bbold}
\usepackage{float}
\usepackage{subfloat}
\usepackage[squaren]{SIunits}
\usepackage{xcolor}

\usepackage[caption=false]{subfig}
\usepackage{tikz}
\usetikzlibrary{arrows.meta} 
\usepackage{makecell}
\usepackage{subfig}
\usepackage{pifont}   
\usepackage{graphicx} 
\graphicspath{{Figures/}}
\usepackage{dcolumn}  
\usepackage{bm}       
\usepackage{multirow} 
\usepackage{placeins}
\usepackage[colorlinks]{hyperref}
\usepackage{mathtools}
\usepackage{appendix}

\begin{document}
	
	\title{Anisotropic tunneling through magnetic barriers in 
		8‑Pmmn borophene}
		

	\author{Rachid El Aitouni}
	\affiliation{Laboratory of Theoretical Physics, Faculty of Sciences, Choua\"ib Doukkali University, PO Box 20, 24000 El Jadida, Morocco}
	
	\author{Sanae Zriouel}
	\affiliation{Cadi Ayyad University, UCA, Faculty of Sciences and Technologies, IMED-Lab, Study Group of Optoelectronic Materials, Marrakech, Morocco.}
%


 \author{Clarence Cortes}
 \affiliation{Vicerrector\'ia de Investigaci\'on y Postgrado, Universidad de La Serena, La Serena 1700000, Chile}
 \author{David Laroze}
 \affiliation{Instituto de Alta Investigación, Universidad de Tarapacá, Casilla 7D, Arica, Chile}

	\author{Ahmed Jellal}
	\email{a.jellal@ucd.ac.ma}
	\affiliation{Laboratory of Theoretical Physics, Faculty of Sciences, Choua\"ib Doukkali University, PO Box 20, 24000 El Jadida, Morocco}

	\begin{abstract}
We present a theoretical study of electron tunneling through a magnetic barrier in 8-Pmmn borophene, created by depositing two ferromagnetic strips on the borophene sheet. Using a low-energy effective Hamiltonian that captures the anisotropic Dirac spectrum, we solve the Dirac equation in three regions and impose wave-function continuity at the interfaces. From the resulting spinor solutions, we compute current densities and determine transmission and reflection probabilities as functions of incident energy, angle, and barrier parameters. The transmission exhibits strong anisotropy due to the tilted Dirac cones, with pronounced suppression for specific incident directions, suggesting directional filtering of carriers. We further calculate the conductance using the Landauer-Büttiker formalism, revealing that both magnetic strength and barrier width can tune  {the charge transport properties. The results demonstrate that} engineered magnetic barriers in 8-Pmmn borophene enable precise control over electron flow, offering a platform for {anisotropic transport control} and tunable quantum devices. The interplay between the intrinsic anisotropy of borophene and external magnetic barriers provides rich opportunities to manipulate Dirac fermions in two-dimensional systems.
	\end{abstract}
	
	\pacs{78.67.Wj, 05.40.-a, 05.60.-k, 72.80.Vp\\
		{\sc Keywords}: Borophene, magnetic barrier, Dirac equation, transmission, conductance.}
	\maketitle

	\section{Introduction}	\label{Intro}

    The achievements of isolating graphene, 
    {which consists of a single atomic layer of carbon arranged in a hexagonal lattice,}
    brought about a vital change in the field of materials science 
    {was recognized by the award of the}
    \cite{novoselov2004, Novoselov2005two}. Through its outstanding electrical and mechanical and thermal characteristics 2D material enables multiple technological applications \cite{neto2009electronic,silvestre2015folded,ji2012atomic}. 
    {Inspired by the remarkable performance of graphene-based nanomaterials, researchers have explored borophene, a two-dimensional material composed of a single layer of boron atoms that can form various honeycomb-like structures} \cite{piazza2014planar,li20202d,kaneti2021borophene,zhang2017two}.
Theoretical studies have proposed several structural phases that exist in both bulk and 2D boron allotropes which include the $\alpha$ and $\beta$ allotropes and other phases \cite{Zhang2018O,kong2021oblique,lopez2016electronic,gonzalez2008boron,piazza2014planar}. 
{Because of the complex bonding characteristics of boron, the formation of a stable pure honeycomb lattice is challenging. However, stable planar structures can be achieved through a combination of hexagonal and triangular motifs \cite{tang2007novel,tang2009self}. The $2B$:$Pmmn$ phase consists of two boron atoms in its primitive unit cell and crystallizes in an orthorhombic structure described by the space group 59 ($Pmmn$).
	Moreover, hydrogenated borophene (commonly referred to as borophane) has been proposed as a promising Dirac material, exhibiting Dirac-like electronic properties and a Fermi velocity that can approach nearly twice that of graphene.}

The distinctive properties of borophene establish it as 
{a promising two-dimensional material with characteristics that differ significantly from those of graphene.
The continuous exploration of graphene and borophene provides valuable insights, as these materials exhibit complementary properties with important implications} in
nanoscale physics and materials science \cite{jugovac2023coupling, liu2019borophene, hou2021borophene,  mortazavi2020machine}.
The study of 2D materials requires careful investigation of their interfaces {and heterostructures, where the interaction between different materials can lead to complex physical behaviors and emergent phenomena \cite{das2015beyond}. These interfaces are particularly interesting because the combination of distinct material properties can generate functionalities that are absent in the individual components.
In this context, graphene–borophene hybrid structures represent a promising research direction. Their interfaces are expected to exhibit novel electronic, structural, and transport properties that extend beyond those of isolated graphene or borophene \cite{xie2021chemistry}. The investigation of such heterostructures provides fundamental insight into interfacial interactions and offers potential opportunities for designing next-generation nanoscale electronic devices \cite{ma2020review, usman2021bismuth, riazi2021ti}.
Recent studies have demonstrated the importance of these concepts through investigations of graphene-based junctions \cite{iqbal2023nanostructures, PhysRevB.108.245419, behura2019graphene} and borophene-based structures \cite{khalatbari2021spin, xu2023valley}, which continue to attract significant interest for future developments in quantum transport and advanced material engineering.}

    In addition to electrostatic and mass-induced confinement, magnetic barriers provide a highly versatile and direct method to manipulate electron transport in two-dimensional Dirac materials. In 8-Pmmn borophene, the intrinsic anisotropy and tilting of the Dirac cones lead to strongly 
    {strongly direction-dependent tunneling properties}, making magnetic barriers particularly effective for controlling carrier trajectories. When a barrier is engineered using ferromagnetic strips, it generates a 
    {localized magnetic vector-potential barrier that modifies the electronic states and scattering processes.}
    The resulting tunneling characteristics are highly sensitive to incident energy, angle, and barrier geometry, leading to pronounced anisotropic transmission and the possibility of directional 
    {control and filtering of charge carriers.}
   Beyond single-particle tunneling, the conductance of such structures, evaluated via the Landauer-Büttiker formalism, exhibits tunable features that depend on both the magnetic barrier parameters and the intrinsic electronic structure, allowing for precise modulation of {the charge transport response}. Moreover, interference effects and quasi-bound states within the barrier region can give rise to resonances and robust transport signatures, providing additional handles to engineer electron flow. These combined features position magnetic-barrier-engineered 8-Pmmn borophene as a promising platform for {anisotropic transport control and tunable quantum devices}, bridging the gap between fundamental studies of Dirac fermions and practical applications in tunable, low-dimensional electronic systems.
   

We present a comprehensive theoretical investigation of electron tunneling through a magnetic vector-potential barrier in 8-Pmmn borophene, realized by depositing two ferromagnetic strips on the borophene sheet. The barrier is generated by two oppositely oriented delta-function magnetic fields at the interfaces, which produce a constant vector potential in the central region. This configuration, combined with the intrinsic tilted and anisotropic Dirac cone spectrum of 8-Pmmn borophene, gives rise to a rich tunneling behavior that differs significantly from that of isotropic Dirac materials such as graphene. Employing a low-energy effective Hamiltonian, we solve the Dirac equation in the three regions of the system and match the spinor wave functions at the interfaces to obtain analytical expressions for the transmission and reflection probabilities. The resulting transmission shows a strong directional dependence, with pronounced suppression for specific incident angles and the emergence of resonant tunneling features that can be controlled by tuning the barrier width and magnetic strength. To connect these microscopic results to measurable quantities, we evaluate the conductance within the Landauer–Büttiker formalism, revealing highly tunable transport characteristics governed by the interplay between the barrier geometry and the anisotropic band structure. 
These results demonstrate the potential of magnetic barriers in 8-Pmmn borophene for controlling anisotropic carrier transport in two-dimensional quantum devices.

    The paper is organized as follows. In Sec.~\ref{II}, we present the mathematical model of low-energy electrons in 8-Pmmn borophene under a magnetic barrier and derive the analytical solutions for the energy spectrum inside and outside the barrier. Sec.~\ref{III} is devoted to the formulation of transmission and conductance through the continuity of the wave function and current densities. {In Sec.~\ref{IV}, we provide a detailed numerical analysis of the transmission probabilities, conductance, 
    emphasizing the effects of barrier width, magnetic strength, and the intrinsic anisotropy of the Dirac cones}. Sec.~\ref{V} concludes the paper with a summary of the main results and a discussion of their implications {for quantum device and directional-filtering applications.}


\section{Theoretical model}\label{II}
\begin{figure}[ht!]
    \centering
    \includegraphics[scale=0.16]{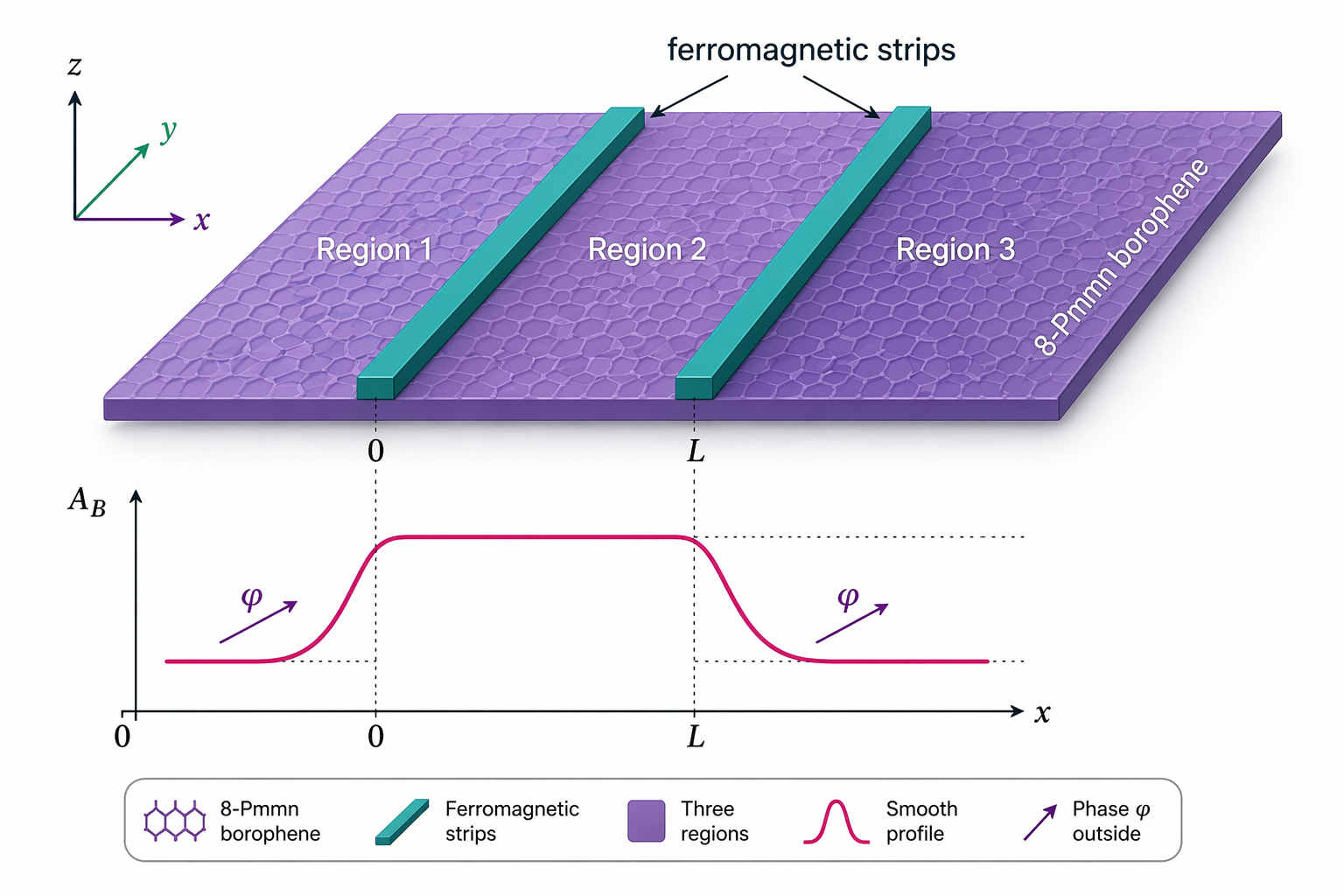}
    \caption{
   The schematic magnetic barrier profile, generated by depositing two ferromagnetic strips on an 8‑Pmmn borophene sheet, creates three distinct regions labeled 1, 2, and 3.}
    \label{str}
\end{figure}

We consider a monolayer sheet of 8-Pmmn borophene divided into three consecutive regions along the (x)-direction, as illustrated in Fig.~\ref{str}. The left \((x<0)\) and right {\((x>L)\)} regions create field-free zones which function as the areas where Dirac fermions enter and exit the system, whereas the intermediate region \((0<x<L)\) is subjected to a perpendicular magnetic field \(B\). This spatially confined magnetic field defines a magnetic barrier of width \(L\), which strongly modifies the carrier dynamics inside the central region. Due to the anisotropic and tilted Dirac cone structure of 8-Pmmn borophene, the response of the quasiparticles to the magnetic barrier is expected to differ significantly from that of isotropic Dirac materials. In particular, the field induces phase accumulation and wave-vector mismatch across the interfaces, leading to reflection, transmission, and the formation of quasi-bound states within the barrier. Such a setup provides a convenient platform for investigating magnetic-field-controlled transport phenomena, including transmission resonances, lateral beam shifts, and delay-time effects in anisotropic Dirac systems.

{The Hamiltonian describing the motion of a spin‑up electron in valley $K$	through this structure is expressed as} \cite{sadhukhan2017, nakhaee2018tight, Zhou2019Valley,zabolotskiy2016strain}
\begin{widetext}
      \begin{align}
   H = v_x p_x \sigma_x + v_y \left[ p_y + {eA_B(x)} \right]\sigma_y  +  v_t \left[ p_y + {eA_B(x)} \right]\sigma_0
\end{align}
\end{widetext}
where $v_x=0.86\times 10^6$ m/s $v_x$ and $v_y=0.46\times 10^6$ m/s are the anisotropic Fermi velocities, while $v_t=0.32 \times 10^6$ m/s) is the tilt velocity, $\sigma_{x,y}$ are the Pauli matrices, $\sigma_0$ is the identity matrix,  ($p_x,p_y$) are the momentum operators, with  $p_j = -i \hbar \partial_j$, and 
{$A_B(x)$ denotes the vector potential associated with the magnetic barrier, created by two oppositely oriented delta-function magnetic fields at the interfaces $x=0$ and $x=L$, generated by two ferromagnetic strips placed on the 8-Pmmn borophene layer}. We consider these strips to be infinite along the $y$-direction in order to neglect edge effects. 
The magnetic field takes the form
\begin{align}
    B(x) = B \left[\delta(x) - \delta(x-L) \right]
\end{align}
where $\delta$ is the Dirac delta function. 
{It should be noted that this field is concentrated at the two interfaces as delta-function spikes, while the central region supports a constant vector potential, constituting a magnetic vector-potential barrier rather than a uniform magnetic field inside the barrier region.}
Consequently, $A_B(x)$ can be written as
	\begin{equation}
		A_B(x)=Bl_B\left[\Theta(x)-\Theta(x-L)	\right]
		\end{equation}
		with  $l_B=\sqrt{\frac{\hbar}{eB}}$ is the magnetic length
		and $\Theta(x)$ denotes the step function.

To determine the solutions for the energy spectrum, we consider the
the eigenvalue equation to obtain
\begin{align}
     &(E-v_t\Pi_y)\psi_A(x,y)=(v_xp_x-iv_y\Pi_y)\psi_B(x,y)\\
       &(E-v_t\Pi_y)\psi_B(x,y)=(v_xp_x+iv_y\Pi_y)\psi_A(x,y)
\end{align}
where  $\Pi_y = p_y + e A_B(x)$ is the conjugate momentum.
The motion of fermions is considered along the $x$-axis, from left to right. Thus, the spinors in the three regions of our structure take the form $\Psi(x,y) = (\psi_A(x), \psi_B(x))^T e^{ik_yy}$, with $T$ denotes the transpose. By using the dimensionless parameters
$E_0=\frac{\hbar v_t}{l_B}$, $x=x/l_B$, $k_y=k_yl_B$, $v_x=v_x/v_t$, $v_y=v_y/v_t$, and $\epsilon=E/E_0$,  we get
\begin{align}
&(\epsilon- (k_y+1))\psi_A(x)=-i\left(v_x\partial_x+v_y (k_y+1)\right)\psi_B(x)\label{psiA}\\
&(\epsilon- (k_y+1))\psi_B(x)=-i\left(v_x\partial_x-v_y (k_y+1)\right)\psi_A(x)\label{psiBB}
\end{align}
{To decouple the two components, we first solve~(\ref{psiBB}) for $\psi_B(x)$ 
in terms of $\psi_A(x)$
\begin{equation}\label{psiB2}
	\psi_B(x) = \frac{-i}{(\epsilon - (k_y+1))}\left(v_x\partial_x - v_y(k_y+1)\right)\psi_A(x).
\end{equation}
Substituting this expression into~(\ref{psiA}), we obtain a second-order 
differential equation for $\psi_A(x)$ alone
\begin{equation}\label{psiB22}
	(\epsilon - (k_y+1))^2\psi_A(x) = \left(-v_x^2\partial_x^2 + v_y^2(k_y+1)^2\right)\psi_A(x)
\end{equation}
Rearranging, this gives}
\begin{align}
	\partial_x^2 \psi_A(x)+ q_x^2 \psi_A(x) =0
\end{align}
where $q_x$ is defined as
%
\begin{align}
   q_x=\frac{1}{v_x}\sqrt{\left(\epsilon -(k_y+1)\right)^2 -\left( v_y\right)^2 (k_y+1)^2}.
\end{align}
We have obtained a simple differential equation to solve, which admits as solution the plane wave
\begin{align}
    \psi_A(x) = c_1 e^{i k_x\, x} + c_2 e^{-i k_x\, x}
\end{align}
Now, to determine the other component $\psi_B$ of the wave function, we use  \ref{psiB2} to obtain
\begin{align}
    \psi_B(x) =  c_1 \beta e^{i k_x x} - c_2 \beta^* e^{-i k_x x}
\end{align}
with the complex number
	$\beta$ and angle $\theta$ are 
	\begin{align}
	&\beta =\frac{v_xq_x +i v_y (k_y+1)}{(\epsilon - (k_y+1))}=se^{i\theta}\\
&
	\theta=\arctan\!\left(\frac{k_y+1}{q_x}\right).
\end{align}
Finally, the eigenspinor in the intermediate region 2 where the field is applied is written as
\begin{align}
    \Psi_2(x,y)=\left[
   c_1 \begin{pmatrix}
        1\\ \beta
    \end{pmatrix}e^{ik_xx}+c_2
    \begin{pmatrix}
        1\\ -\beta^*
    \end{pmatrix}e^{-ik_xx}
    \right]e^{ik_yy}
    \end{align}
and  the corresponding eigenvalue is
    \begin{align}\label{eq15}
        E = (k_y+1) \;\pm\; \sqrt{(v_x  k_x)^2 + (v_y (k_y+1))^2}
    \end{align}
In regions 1 and 3,  the eigenspinors are determined by solving the eigenvalue equations {in the absence of } the applied field. 
{In this case, we recover the same solution as in~\eqref{psiB22}, with the replacement $k_y+1 \rightarrow k_y$. As a result, we obtain}
\begin{align}
  &  \Psi_1(x,y)=\left[
    \begin{pmatrix}
        1\\ \gamma
    \end{pmatrix}e^{ik_xx}+r
    \begin{pmatrix}
        1\\ -\gamma^*
    \end{pmatrix}e^{-ik_xx}
    \right]e^{ik_yy}
    \\
   & \Psi_3(x,y)=
   t \begin{pmatrix}
        1\\ \gamma
    \end{pmatrix}e^{ik_xx}e^{ik_yy}
    \end{align}
where $r$ and $t$ denote the reflection and transmission coefficients,
	respectively. The complex number $\gamma$ and $k_x$ have the forms
	\begin{align}
		&\gamma=\frac{v_xk_x + i v_y k_y}{\epsilon - k_y}=e^{i\varphi}\\
		& k_x=\frac{1}{v_x}\sqrt{ (\epsilon - k_y)^2 - v_y^2 k_y^2}
	\end{align}
	so that the incidence angle is given by
	\begin{equation}
		\tan\varphi=\frac{v_y k_y}{v_x k_x}.
		\label{eq30}
	\end{equation}
	This differs from the wave-vector angle $\arctan(k_y/k_x)$ due to
	anisotropy ($v_x\neq v_y$). The group velocity
	$\mathbf{v}_g=\nabla_{\mathbf{k}}E(\mathbf{k})$ is not parallel to
	$\mathbf{k}=(k_x,k_y)$ in the tilted-cone system, with components
	\begin{align}
		& v_{g,x}=\frac{v_x^2 k_x}{\sqrt{(v_x k_x)^2+(v_y k_y)^2}}\\
		& v_{g,y}=v_t+\frac{v_y^2 k_y}{\sqrt{(v_x k_x)^2+(v_y k_y)^2}}.
	\end{align}
	Thus $\varphi$ also differs from the group-velocity direction due to
	tilt ($v_t\neq 0$). The angle $\varphi$ is the relevant quantity for
	the matching conditions and current densities because it enters
	directly through the spinor structure.


\section{Transmission and Conductance}\label{III}
{In order to disregard edge effects of the ferromagnetic strips, their length is assumed to be infinite along the $y$-axis, i.e., much larger than the barrier width \cite{infini1,infini2}.}
To compute the transmission and conductance, we math the eigenspinors at the interfaces $x=0$ and $x=L$.  Then from $\Psi_1(0,y)=\Psi_2(0,y)$ and $\Psi_2(L,y)=\Psi_3(L,y)$,  we obtain
   \begin{align}
    &1+r=c_1+c_2\\
    &s'\,e^{i\varphi}-r\,s'\,e^{-i\varphi}=c_1\,s\,e^{i\theta} - c_2s\,e^{-i\theta}a_2\\
    &c_1\,e^{iq_xL}+c_2\,e^{-iq_xL}=t\,e^{ik_xL}\\
    &c_1\,s\,e^{i\theta} e^{iq_xL}-c_2\,s'\,e^{-i\theta}e^{-iq_xL}=t\,s'\,e^{i\varphi} e^{ik_xL}.
\end{align}
This set can be solved to get the reflection and transmission coefficients
\begin{widetext}
    \begin{align}
	r&=\dfrac{\sin(q_x L)(\sin\varphi-\sin\theta)e^{i\varphi}}
	{e^{2 i \varphi} (e^{2 i q_x L} + e^{2i\theta}) + 1 + e^{2i(q_x L + \theta)} - 2 e^{i(\theta+\varphi)}(-1 + e^{2 i q_x L}) }\\
	t& = \dfrac{e^{-i L k_x } \cos\theta \cos\varphi}{\cos(q_x L) \cos\theta \cos\varphi - i \, \sin(q_x L) (1 - \sin\theta \sin\varphi)}.
\end{align}
\end{widetext}

The transmission probability can be evaluated from the ratio between the transmitted and incident
current densities. These densities are obtained from the continuity equation
\begin{align}
    \partial_t \rho + \nabla \cdot J = 0
\end{align}
with $\rho = |\psi|^2$. As a result we obtain  the following expressions for the incident, reflected, and transmitted densities, respectively
\begin{align}
   & J_{\text{inc}}= 2 v_x \cos(\varphi)\label{Jinc}\\
  & J_{\text{ref}}= 2 v_x \cos(\varphi) |r|^2\\
   &J_{\text{tra}}= 2 v_x \cos(\varphi) |t|^2.\label{Jtra}
\end{align}
By definition, the transmission and reflection probabilities can be obtained from  $
T = \frac{J_{\text{tra}}}{J_{\text{inc}}}=\left| t \right|^2$ and $R = \frac{J_{\text{ref}}}{J_{\text{inc}}}=\left| r \right|^2$.
After straightforward algebra, we get
   \begin{align}\label{T}
T=\frac{\cos^{2}\theta\,\cos^{2}\varphi}
{\cos^{2}\theta\,\cos^{2}\varphi\,\cos^{2}(q_x L)
+ \sin^{2}(q_x L)\left(1-\sin\theta\,\sin\varphi\right)^{2}}.
\end{align}
	Because $\varphi$ determines the probability flux through the spinor
	structure, the transmission formula \eqref{T} satisfies the
	conservation relation $T+R=1$.

Transmission is intrinsically connected to the microscopic behavior of fermions.
To properly understand their dynamics in 8-Pmmn borophene, the conductance is
calculated from the transmission probability, thereby linking microscopic processes to macroscopic observables. According to the Landauer-Büttiker formalism \cite{butker, conduct1}, the conductance is expressed as:
\begin{align}
		G &= G_0 \int_{-k_y^{\text{max}}}^{k_y^{\text{max}}} T(E,k_y) dk_y
	\end{align}
	where $G_0$ denotes the conductance unit, $E_F$ is the Fermi energy,  and $k_y^{\text{max}}$ represents the maximum wave vector component along the $y$-direction. Taking into account that $k_y^{\text{max}} = k \sin\varphi^{\text{max}}$, the conductance  can be expressed as
	\begin{align}
		G = G_0 \int_{-\varphi^{\text{max}}}^{\varphi^{\text{max}}} T(E, \varphi) \cos\varphi \, d\varphi.
	\end{align}
%
{The highest value of $\phi_{\max}$ is determined by the condition that $k_x$ is real. From~(19), we obtain $({\epsilon} - k_y)^2 \geq v_y^2 k_y^2$, from which $k_y^{\max} = \epsilon/(1+v_y)$ and $\phi_{\max} = \arcsin(k_y^{\max}/k)$. Modes beyond this threshold are evanescent and carry no current.
The anisotropy of 8-Pmmn borophene affects the density of transverse modes. Since $v_x \neq v_y$, the Fermi contour is no longer circular, which modifies the distribution of propagating $k_y$ modes compared to isotropic Dirac materials. The integral over $k_y$ in~(32) accounts for this naturally by sampling all propagating modes directly.
	The $\cos\phi$ factor in~(33) arises from the Jacobian of the change of variables from $k_y$ to $\phi$. Because the tilt velocity $v_t$ shifts the group velocity away from the wave vector direction, $\phi$ is measured with respect to the group velocity rather than $\mathbf{k}=(k_x,k_y)$, which ensures consistency with the probability flux. The $\cos\phi$ weighting thus correctly reflects the projected current in the tilted cone geometry.}
	Alternatively, the conductance can be interpreted as the average flux of fermions across half of
	the Fermi surface \cite{Biswas}
	

\section{Numerical results}\label{IV}

To investigate the transport properties of Dirac fermions in 8-Pmmn borophene under a magnetic barrier, we systematically analyze the transmission and the corresponding conductance as functions of incident angle, energy, transverse momentum, and barrier width. These quantities provide complementary insights into the microscopic and macroscopic aspects of electron transport. In particular, the transmission captures the underlying quantum scattering processes, while the conductance reflects the cumulative contribution of all propagating modes.

\begin{figure}[ht!]
	\centering
	\subfloat[$\epsilon=1$]{\includegraphics[scale=0.5]{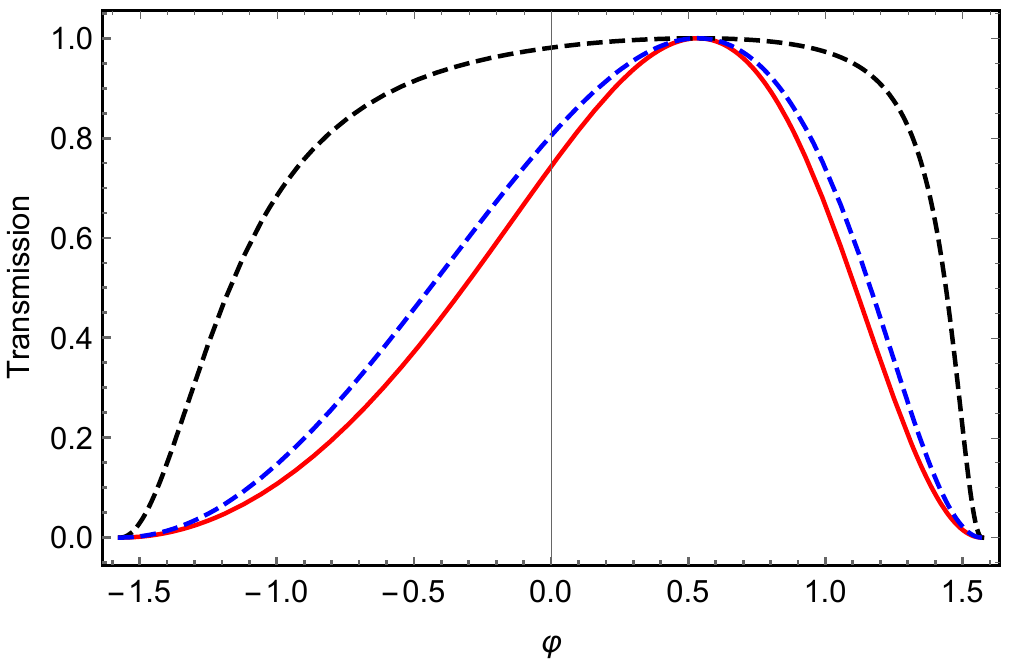}\label{Tpa}}\\
	\subfloat[$\epsilon=3$]{\includegraphics[scale=0.5]{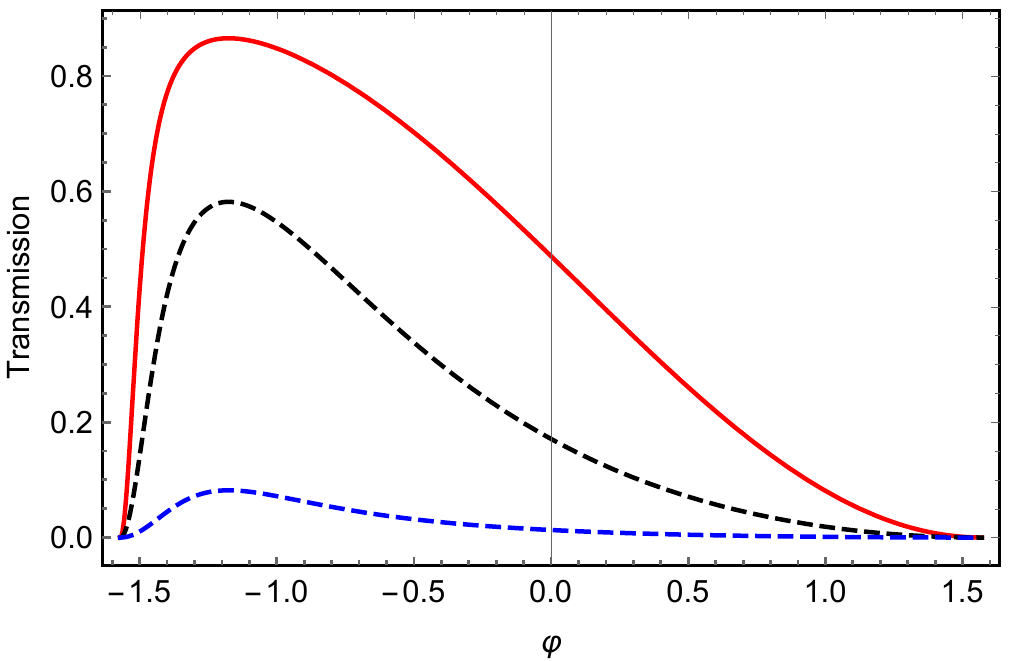}\label{Tpb}}\\
	\subfloat[$\epsilon=10$]{\includegraphics[scale=0.5]{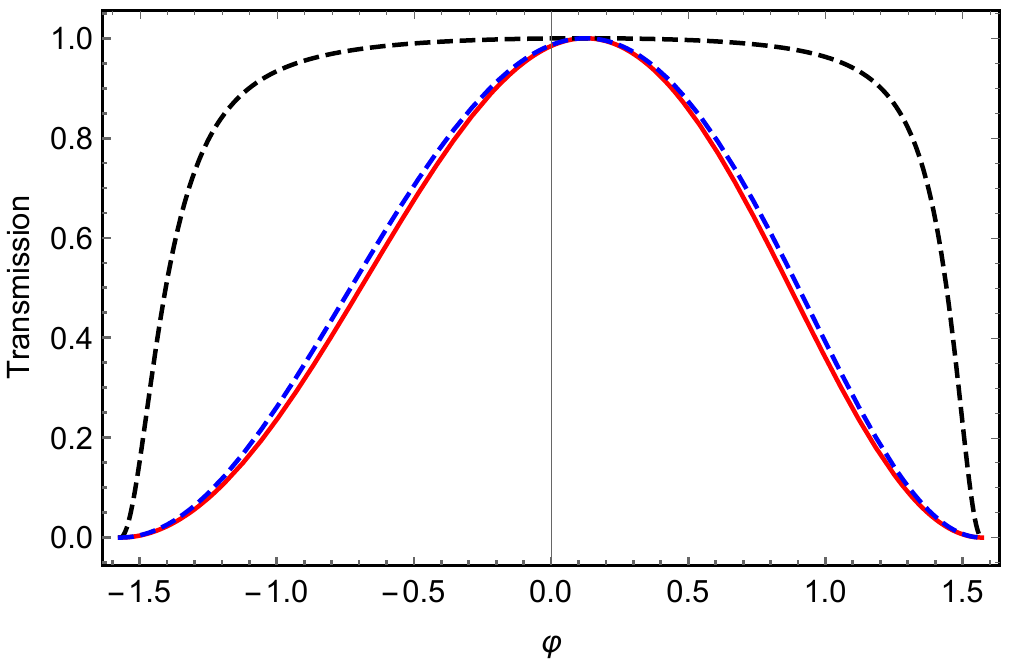}\label{Tpc}}
	\caption{Transmission  as a function of the incident angle $\varphi$ for the transverse momentum $k_y\,l_B=1$, 
		three incident energies (a): $\epsilon=1$, (b): $\epsilon=3$, (c): $\epsilon=10$,
		and three barrier widths $L=1l_B$ (red curve), $2l_B,$ (black curve) and $5l_B$ (blue curve).}\label{Tp}
\end{figure}

The transmission probability as a function of the incident angle $\varphi$ for three different energies and barrier widths is shown in Figure~\ref{Tp}.
In Fig.~\ref{Tpa} at low energy ($\epsilon = 1$), the transmission is substantially lowered over most angular ranges, especially as the barrier width grows. The narrow transmission peaks shifted away from normal incidence suggest that only a small number of incident directions help in tunneling. This behavior is caused by the predominance of evanescent modes inside the magnetic barrier and shows the great anisotropy of the Dirac spectrum in 8-Pmmn borophene.
The transmission rises noticeably and displays larger angular windows at intermediate energy ($\epsilon = 3$) in Fig.~\ref{Tpb}. Particularly for greater barrier widths, oscillatory properties start to show, which can be explained by quantum interference effects resulting from several reflections within the barrier area. The angular asymmetry stays, therefore confirming that the slanted Dirac cones move the conditions for best transmission away from regular incidence.
In Fig.~\ref{Tpc}, at high energy ($\epsilon = 10$), the transmission approaches unity over a broad angular range, showing that dominating modes of propagation govern transportation. Though little oscillations owing to phase accumulation within the barrier persist, reliance on barrier width fades. The angular asymmetry is still apparent but less marked than in the low-energy region.
These data show that the magnetic barrier serves as a directional filter, choosing certain incident angles depending on carrier energy and barrier breadth. This behavior stands in contrast to graphene, in which Klein tunneling results in perfect transmission at normal incidence and instead mirrors the effect of anisotropic and tilted Dirac cones in borophene~\cite{Zhang2018O, sadhukhan2017}.

\begin{figure}[ht!]
	\centering
\subfloat[$k_y\, l_B=1$]{\includegraphics[scale=0.5]{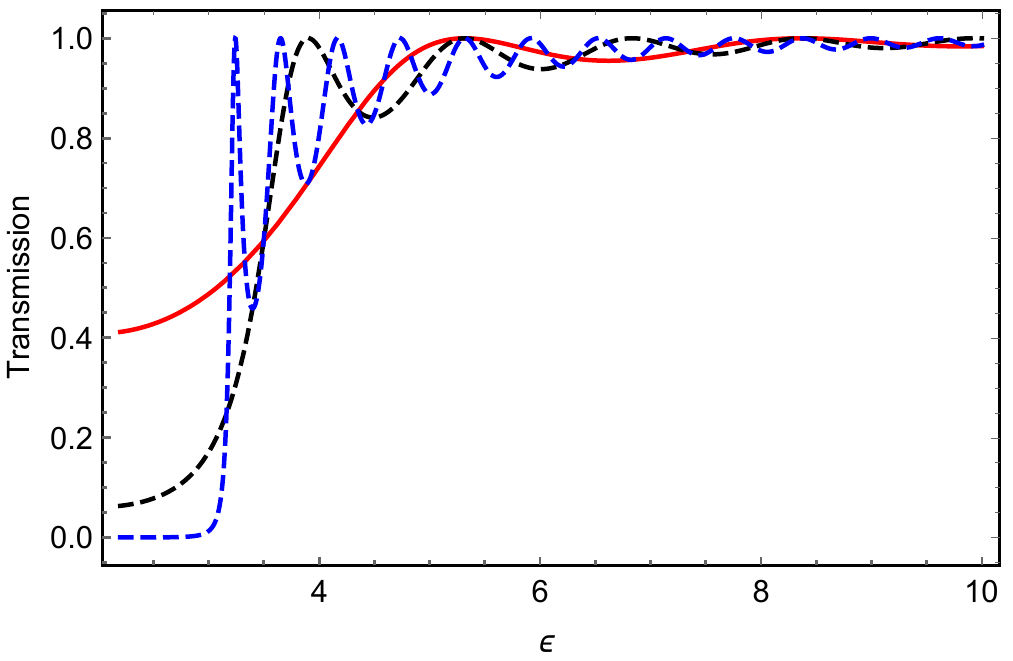}\label{TEa}}\\
	\subfloat[$k_y\, l_B=2$]{\includegraphics[scale=0.5]{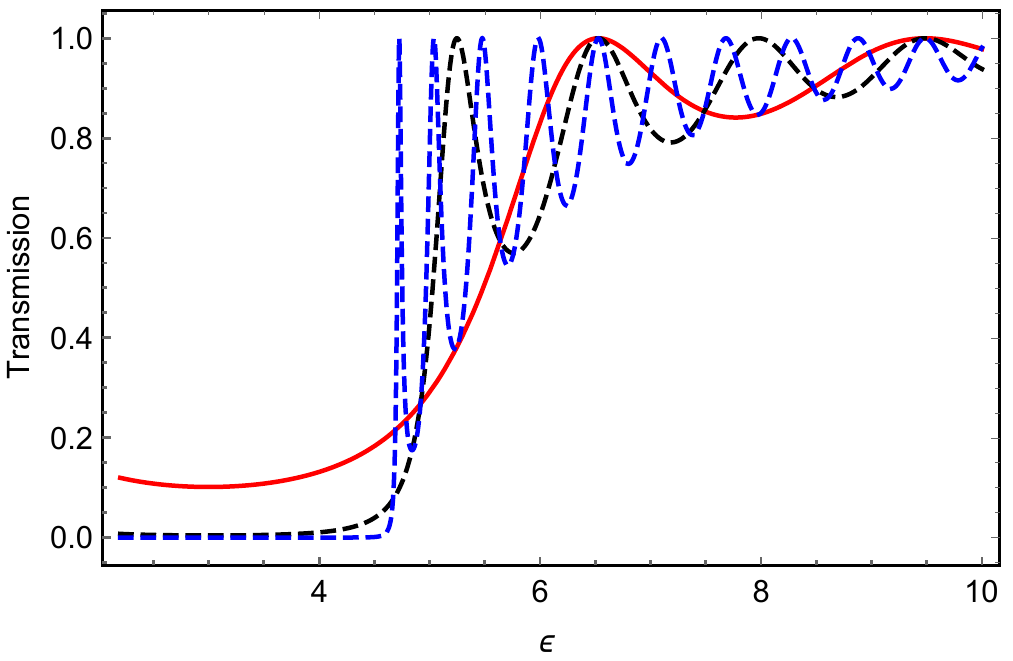}\label{TEb}}\\
	\subfloat[$k_y\, l_B=4$]{\includegraphics[scale=0.5]{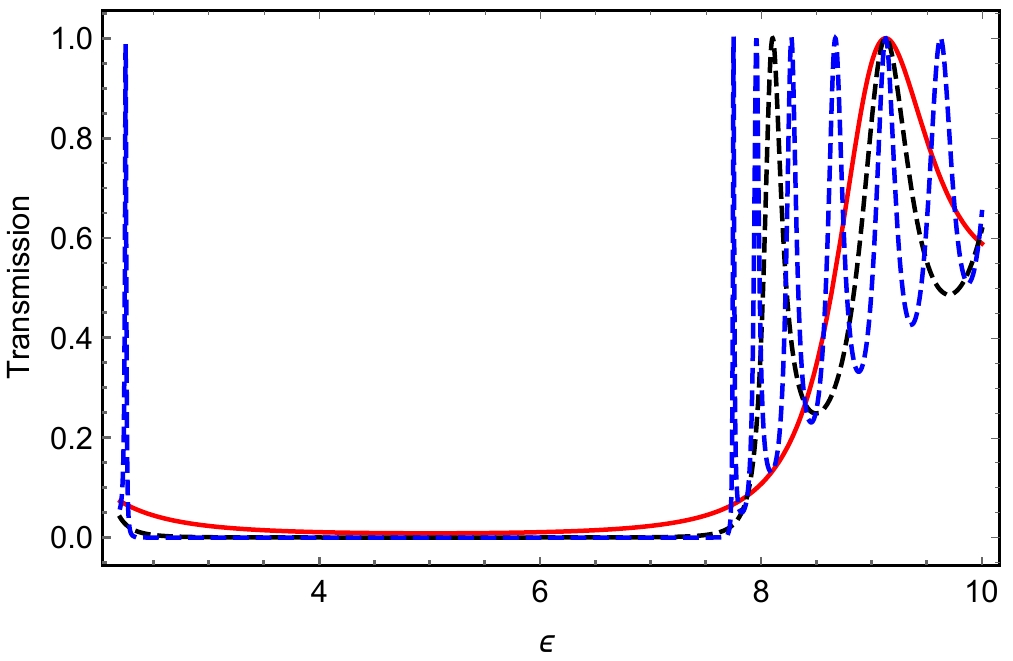}\label{TEc}}
	\caption{Transmission  as a function of the incident energy $\epsilon$ at normal incidence $(\varphi=0)$ for three transverse momenta (a): $k_y\, l_B=1$, (b): $k_y\, l_B=2$, (c): $k_y\, l_B=4$, and three barrier widths $L=1l_B$ (red curve), $2l_B,$ (black curve),  $5l_B$ (blue curves).}\label{TE}
\end{figure}

Figure~\ref{TE}  displays the transmission versus the incident energy $\epsilon$ for different barrier widths and transverse momentum $k_y l_B$.
The transmission increases continuously with energy for all barrier widths in Fig. \ref{TEa} for relatively little transverse momentum ($k_y l_B = 1$). Transmission at low energies is inhibited by the difference between incident and barrier states. The fact that the transmission nears unity as the energy rises points to dominant transportation by traveling modes. At intermediate energies, when larger barriers result in decreased transmission and the start of oscillatory behavior, the effect of the barrier width is most clear.
The suppression of transmission at low energy increases for moderate transverse momentum ($k_y l_B = 2$) in Fig.~\ref{TEb}. The beginning of major transmission changes toward higher energies, therefore showing the growing influence of transverse momentum in impeding tunneling. Larger barrier widths especially at higher energies, oscillations become more apparent because of amplified quantum interference phenomena inside the barrier area.
Transmission is severely reduced over a broad energy range at high transverse momentum ($k_y l_B = 4$) in Fig.~\ref{TEc}, only reaching significant values at enough high energies. Because the threshold energy needed for transmission rises significantly, large $k_y l_B$ virtually serves as an extra barrier. Oscillatory behavior returns when this threshold is crossed, which is associated to resonant tunneling through the potential barrier.
Fig.~\ref{TE} shows that, taken all, raising transverse momentum lowers transmission and pushes the beginning of propagating modes toward higher energies. This behavior agrees with earlier investigations of Dirac fermions in anisotropic systems, in which the transverse momentum is very important in influencing tunneling efficacy and resonance conditions~\cite{Nakhaee2018}.

\begin{figure}[ht!]
	\centering
\subfloat[$k_y\, l_B=0.5$]{\includegraphics[scale=0.5]{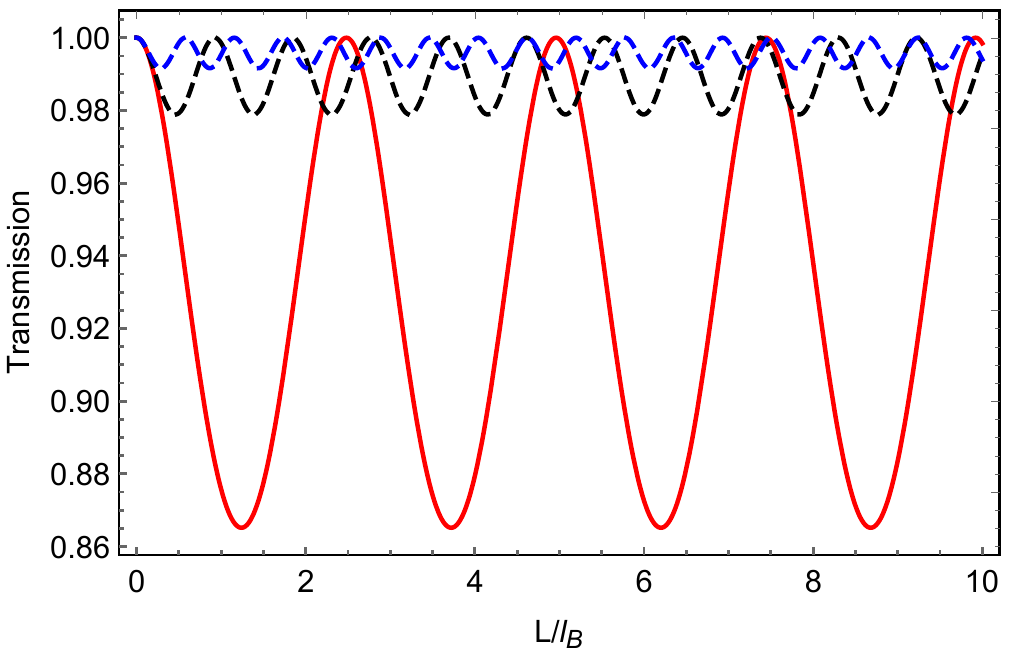}\label{TLa}}\\
	\subfloat[$k_y\, l_B=1$]{\includegraphics[scale=0.5]{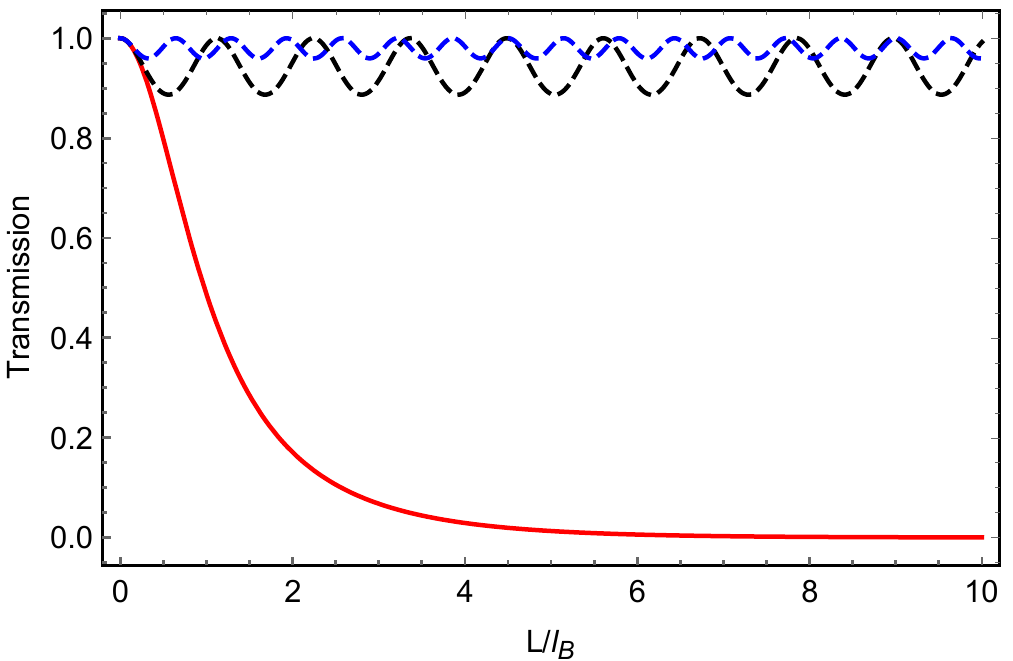}\label{TLb}}\\
	\subfloat[$k_y\, l_B=2$]{\includegraphics[scale=0.5]{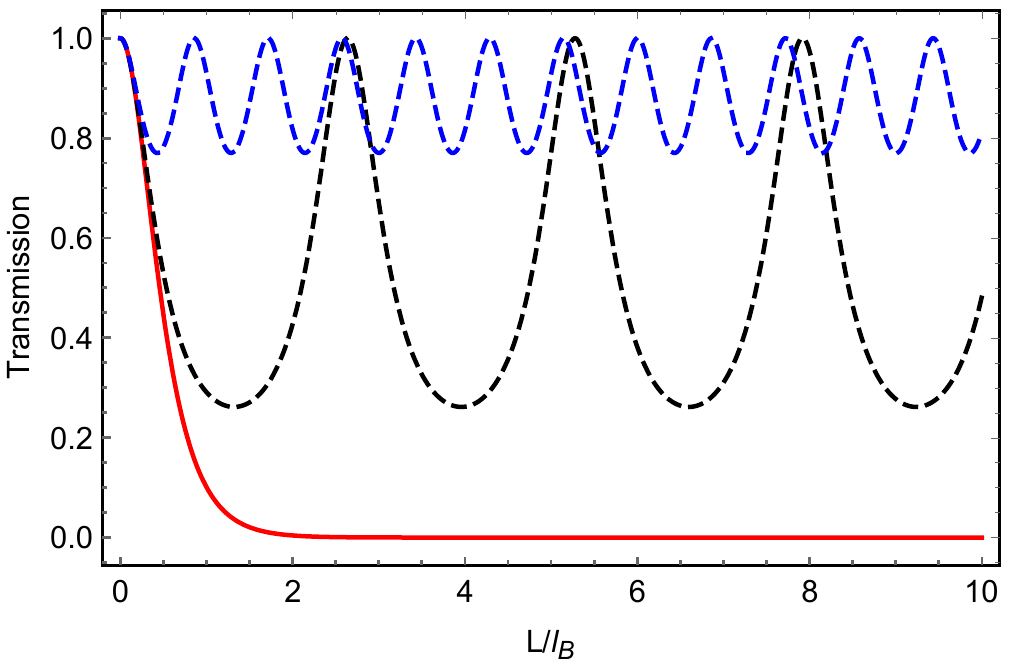}\label{TLc}}
	\caption{Transmission as a function of the normalized barrier width $L/l_B$ at normal incidence $(\varphi=0)$ for three transverse momenta (a): $k_y\, l_B=0.5$, (b): $k_y\, l_B=1$, (c): $k_y\, l_B=2$, and three incident energies $\epsilon=3$ (red curve),  5 (black curve),  $7$ (blue curve).}\label{TL}
\end{figure}

Figure~\ref{TL} presents the transmission  as a function of the barrier width $L/l_B$ for different energies and transverse momenta.
For low transverse momentum ($k_y l_B = 0.5$) in Fig.~\ref{TLa}, the transmission continues near unity over the whole range of barrier widths with only minor fluctuations. This implies that carriers with minimal transverse momentum, corresponding to nearly normal incidence, are almost transparent to the magnetic barrier. While the total high transmission reflects the preponderance of propagating modes, the small oscillations result from phase accumulation within the barrier.
For modest transverse momentum ($k_y l_B = 1$) in Fig.~\ref{TLb}, the transmission becomes more sensitive to the barrier width. It exhibits notable oscillations at higher energies but descends slowly at lower energies as the width widens. Similar to Fabry--Pérot resonances, these oscillations come from quantum interference effects brought on by several reflections within the barrier.
The transmission is significantly suppressed for great transverse momentum ($k_y l_B = 2$) in Fig.~\ref{TLc} as the barrier width increases, especially at low energies. This behavior suggests that evanescent modes control transport, therefore producing an exponential decay of transmission as width increases. Oscillatory behavior shows itself once again at higher energies, therefore indicating the change from evanescent to propagating modes within the barrier.
Fig.~\ref{TL} highlights the crossover between propagating and evanescent regimes governed by both the transverse momentum and the barrier width, a characteristic attribute of tunneling in anisotropic Dirac systems~\cite{nakhaee2018tight}.

\begin{figure}[ht!]
	\centering
\subfloat[$\epsilon=5$]{\includegraphics[scale=0.48]{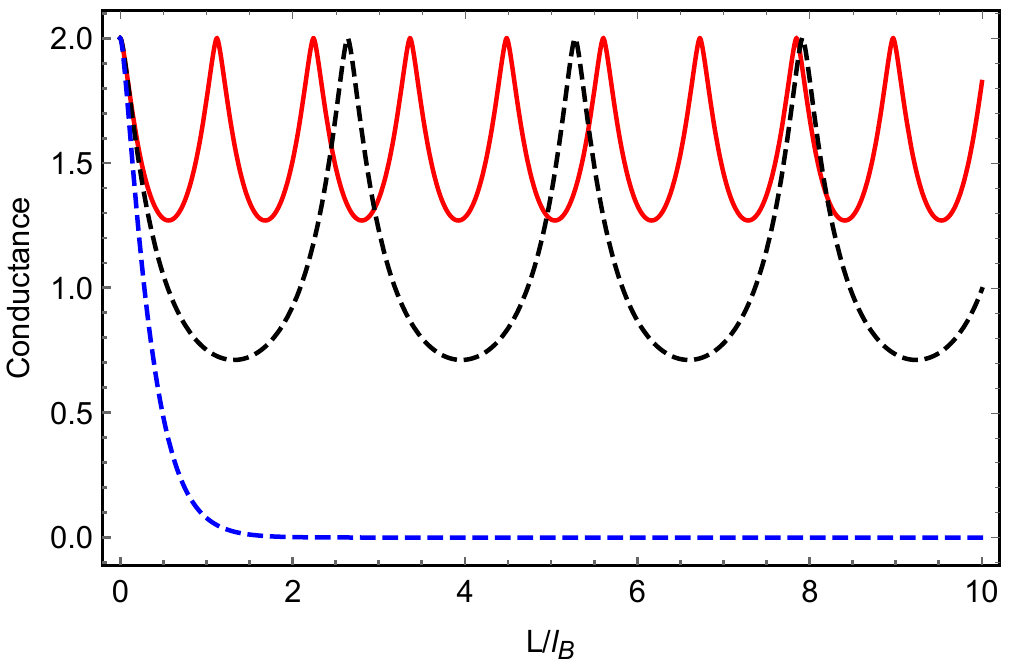}\label{Ga}}\\
	\subfloat[$k_y\,l_B=1$]{\includegraphics[scale=0.48]{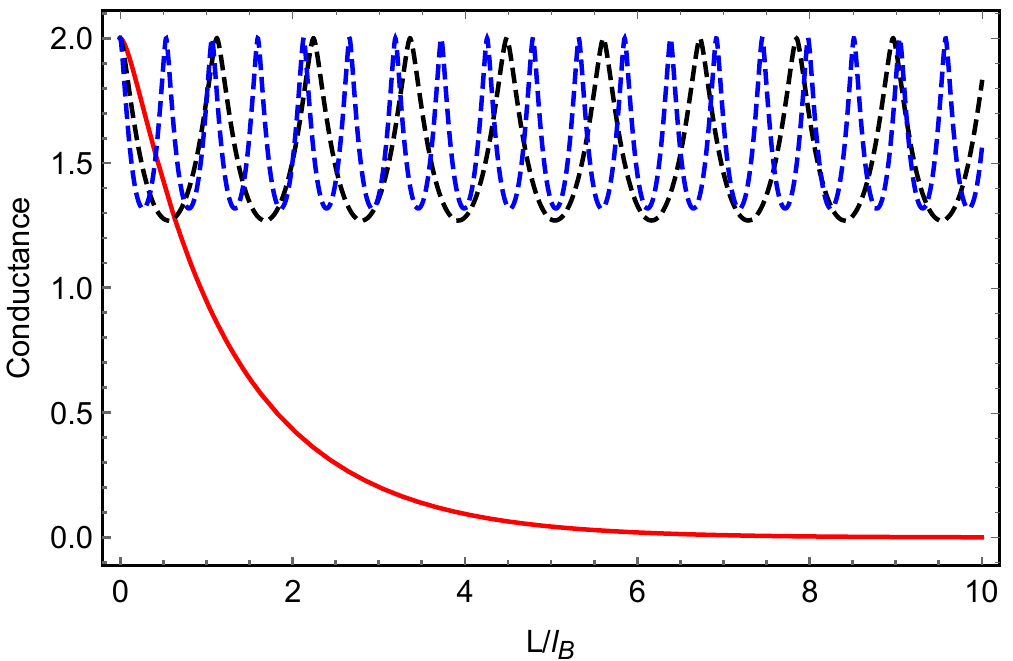}\label{Gb}}\\
	\caption{Conductance as a function of the normalized barrier width $L/l_B$. (a): Incident energy $\epsilon=5$, and  three transverse momenta $k_y\, l_B=1,2,$ and $3$ (red, black, and blue curves, respectively). (b) Results for a fixed transverse momentum $k_y\, l_B=1$ and energies $\epsilon=3,5,$ and $8$ (red, black, and blue curves, respectively).}\label{GL1}
\end{figure}

Figure~\ref{GL1} illustrates the conductance as a function of barrier width $L/l_B$ for several transverse momenta and energies.
The conductance shows various behaviors in Fig.~\ref{Ga} depending on the value of $k_yl_B$ for growing transverse momentum ($k_y l_B = 1, 2, 3$). For small $k_yl_B$, the conductance remains quite high and shows slight oscillations as the barrier width increases, indicating successful barrier passage.   Conductance diminishes as $k_y$ rises, showing the decrease in available propagation modes and the growing influence of evanescent states.
The conductance in Fig.~\ref{Gb} shows oscillatory behavior as a function of the barrier width for various energies ($\epsilon = 3, 5, 8$). These oscillations result from quantum interference phenomena caused by several reflections within the barrier region. Higher energies produce more prominent oscillations and greater conductance values since additional propagation modes add to transportation.
Fig.~\ref{GL1} shows generally that both the barrier width and the transverse momentum have a significant effect on the conductance. This conduct is at odds with graphene, where Klein tunneling, especially at normal incidence~\cite{Nakhaee2018},  keeps conductivity rather strong. But in the current system, the conductance shows a greater suppression at large $k_yl_B $ and more pronounced oscillations with rising energy, thereby reflecting the diminished transmission channels and strengthened interference effects as opposed to graphene.

\begin{figure}[ht!]
	\centering
\subfloat[$L=l_B$]{\includegraphics[scale=0.48]{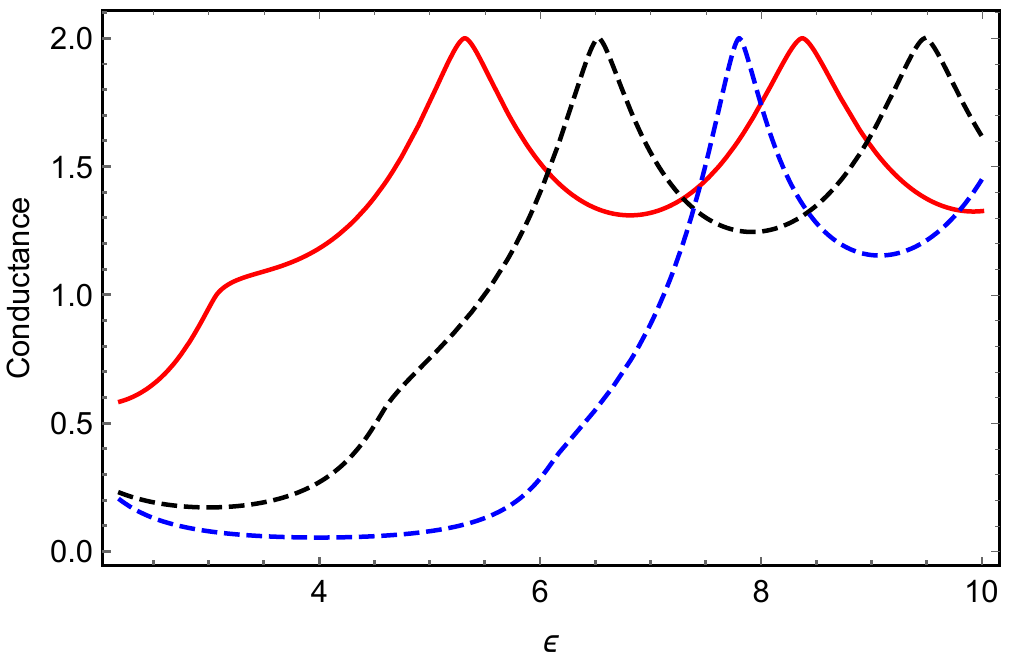}\label{GE5}}\\
	\subfloat[$k_y\,l_B=1$]{\includegraphics[scale=0.48]{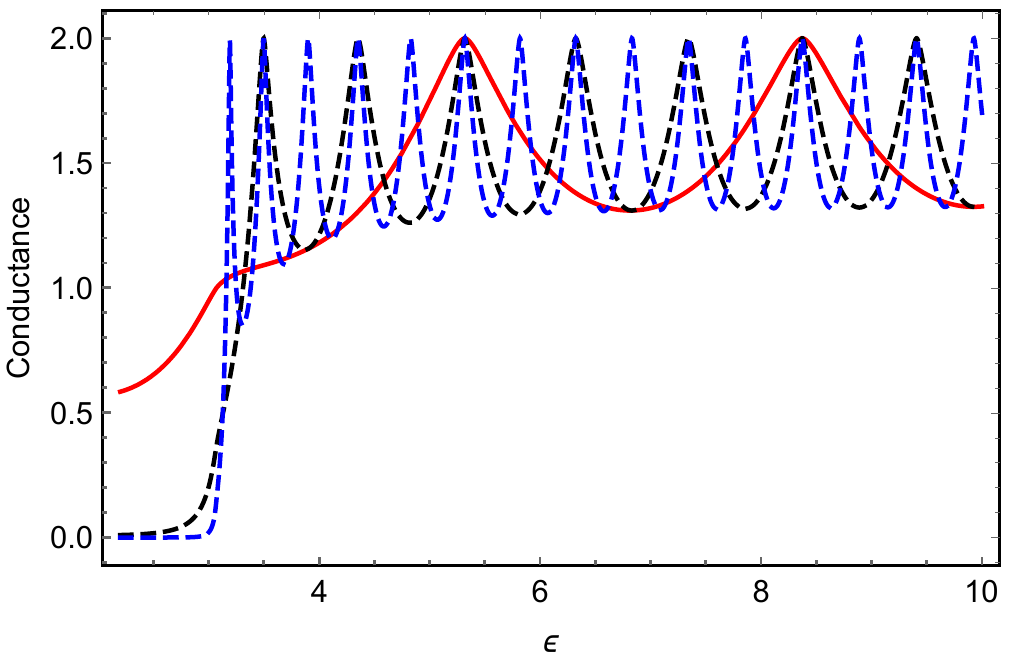}\label{Gk1}}
	\caption{Conductance as a function of the incident energy $\epsilon$. (a) Results for a fixed barrier width $L=l_B$ and transverse momentum $k_y\, l_B=1,2,$ and $3$ (red, black, and blue curves, respectively). (b) Results for a fixed transverse momentum $k_y\,l_B=1$ and barrier widths $L/l_B=1,3,$ and $6$ (red, black, and blue curves, respectively).}\label{GL2}
\end{figure}

Figure~\ref{GL2} displays the conductance vs the incident energy $\epsilon$ for several transverse momenta and barrier widths.
Generally, the conductance rises with energy for rising transverse momentum ($k_y l_B = 1, 2, 3$) in Fig.~\ref{GE5}. Because of the small number of propagating modes, the conductance is suppressed at low energies, especially for large $k_y l_B$. Rising conductance and oscillating behavior reflect the contribution of several transmission channels and interference effects inside the barrier as the energy grows. Higher transverse momentum lowers its total magnitude and pushes the beginning of conductance toward larger energies.
The conductance also rises with energy in Fig.~\ref{Gk1} for various barrier widths ($L/l_B = 1, 3, 6$), but distinct oscillations with a size and frequency depending on the barrier width are seen. More noticeable oscillations resulting from wider widths may be explained by stronger phase accumulation and increased Fabry--Pérot-like interference inside the barrier region.
Fig.~\ref{GL2} shows that the conductance is strongly energy-dependent and regulated by both the transverse momentum and the barrier width. 
These oscillatory features and energy thresholds are characteristic of Dirac-like systems and closely related to interference effects and mode quantization. Comparable effects have been detected in graphene-based materials, where Klein tunneling and Fabry–Pérot resonances are essential elements~\cite{Nakhaee2018,Young2009}.

\begin{figure}[ht!]
	\centering
\subfloat[$L=l_B$]{\includegraphics[scale=0.45]{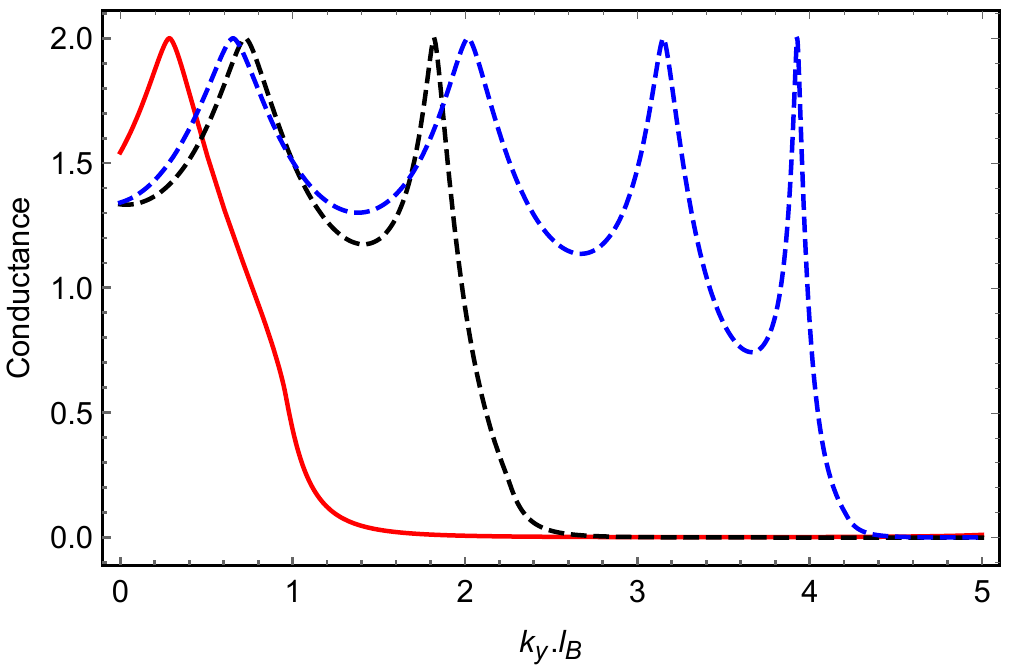}\label{GkE3}}\\
	\subfloat[$\epsilon=5$]{\includegraphics[scale=0.45]{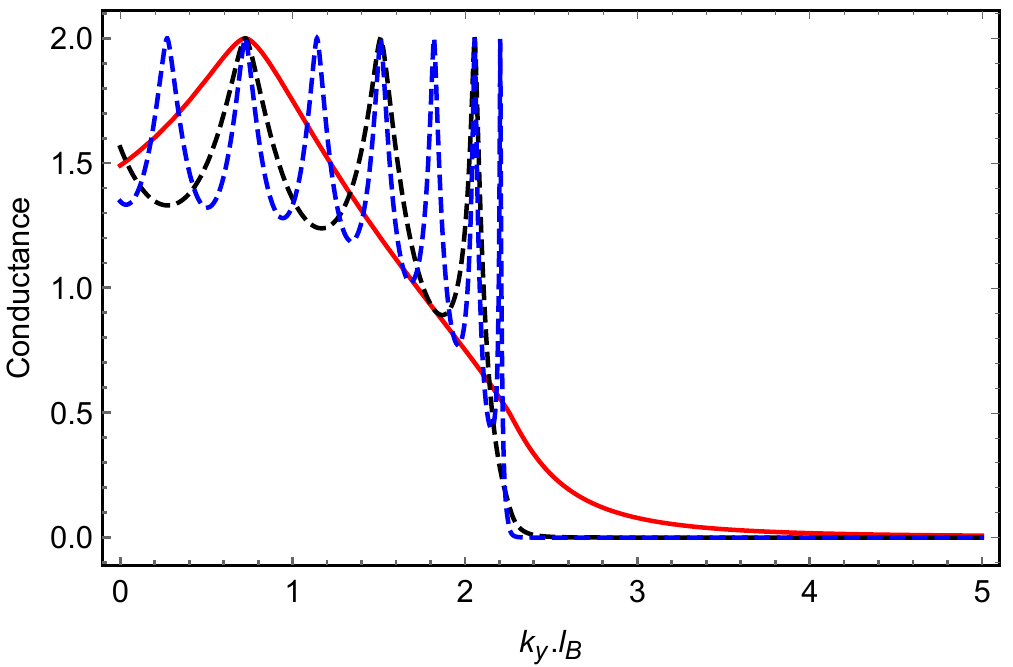}\label{GkE5}}\\
	\caption{Conductance as a function of the transverse momentum $k_y\, l_B$. (a) Results for a fixed barrier width $L=l_B$ and energies $\epsilon=3,5,$ and $8$ (red, black, and blue curves, respectively). (b) Results for a fixed energy $\epsilon=5$ and barrier widths $L/l_B=1,3,$ and $6$ (red, black, and blue curves, respectively).}\label{Gk}
\end{figure}

In Figure~\ref{Gk}, we  show the conductance versus the transverse momentum $k_y l_B$ for different energies and barrier widths. Fig.~\ref{GkE3} indicates that as transverse momentum increases, conductance decreases at energy levels ($\epsilon = 3, 5, 8$).
The conductance is fairly high at small $k_y l_B$, which shows that near-normal incidence carriers significantly aid in transportation. Reflecting the diminishing of propagating modes and the growing dominance of evanescent states, the conductance falls quickly as $k_y l_B$ rises. More transmission channels are present, hence higher energies sustain greater conductance values across a range of $k_yl_B$.
In Fig.~\ref{GkE5}, for several barrier widths ($L/l_B = 1, 3, 6$), the conductance declines with increasing transverse momentum as well; however, the rate of suppression varies with barrier width. Stronger suppression results from bigger widths, therefore showing improved filtering of oblique incident carriers. This conduct emphasizes the magnetic barrier as an angular filter, which permits only a limited range of transverse momenta to contribute notably to transport.
With $k_y l_B$ rising, Fig.~\ref{Gk} shows that the conductance is very sensitive to the transverse momentum, with a distinct change from high-conductance (propagating regime) to low-conductance (evanescent regime). This angular selectivity is a characteristic feature of tunneling in anisotropic Dirac materials and is consistent with earlier research on borophene-based systems~\cite{sadhukhan2017}.

\section{Conclusion}\label{V}

We have conducted a complete theoretical analysis of electron movement through a magnetic barrier in 8-Pmmn borophene. The study demonstrated how Dirac cone intrinsic anisotropy interacts with magnetic fields which researchers applied to the system. We established a method to obtain transmission probabilities and conductance values from current densities by deriving barrier region energy levels through our analytical solution of the low-energy Dirac equation. The results show that tunneling behavior exhibits strong directional dependence because the tilted Dirac cones cause different transmission rates for different incident angles. 
The conductance calculations show that both total conductance 
 can be controlled through barrier width and magnetic strength adjustments. 
The barrier region shows magnetic confinement in 8-Pmmn borophene, which creates quasi-bound states and resonance features that establish new pathways for electron flow control. 
The developed effects create a directional filtering function that selects carriers based on incident angle and transverse momentum, which can be useful for designing tunable quantum transport devices.

Our study establishes magnetic-barrier-engineered 8-Pmmn borophene as a versatile platform for exploring fundamental aspects of Dirac fermion dynamics in low-dimensional systems while also offering practical avenues for device design. The tunable interplay between barrier parameters, anisotropic band structure
 opens the possibility of implementing highly controllable quantum devices with engineered transport characteristics. Finally, the insights gained here not only deepen our understanding of magnetic confinement in anisotropic Dirac materials but also provide a foundation for future experimental investigations, including the design of heterostructures, interference-based devices, and quantum information platforms that exploit the unique properties of 8-Pmmn borophene.

%

\end{document}